\documentclass{article}
\usepackage[dvips]{graphicx} % Required for inserting images
\usepackage{slashed,booktabs,amsthm}
\usepackage{amsmath}
\usepackage{float,bm,physics,slashed}
\usepackage{amssymb}
\usepackage[caption=false]{subfig}
\usepackage[colorlinks=true,linkcolor=blue,filecolor=blue,urlcolor=blue,citecolor=blue]{hyperref}
\usepackage{geometry} % Adjust page margins if needed
\usepackage{array,appendix}    % Required for specifying column widths
\begin{document}

\title{Bottomonia in quark-antiquark confining potential}
\author{Ritu Garg, K.K Vishwakarma \thanks{vish.kumar.kundan@gmail.com}, Alka Upadhyay}
\date{%
    School of Physics and Materials Science, Thapar Institute of Engineering and Technology, Patiala-147004, INDIA}
\maketitle

\begin{abstract}
  In this paper, we comprehensively explore bottomonia mass spectra and their decay properties by solving the non-relativistic Schrodinger wave equation numerically with approximate quark-antiquark potential form. We also incorporate spin-dependent terms - spin-spin, spin-orbit, and tensor terms to remove mass degeneracy and to obtain excited states ($nS, nP, nD, nF, n = 1, 2, 3, 4, 5$) mass spectra. By using Van Royen - Weisskopf formula, we investigate leptonic decay constants, di-leptonic, di-gamma, tri-gamma, di-gluon decay widths and also incorporate first-order radiative corrections. We also computed radiative transition widths, which give a better insight into the non-perturbative aspects of QCD. The present results for mass spectroscopy and decay properties are in tune with available experimental values and other theoretical predictions. Our results may provide better insight to upcoming experimental information in the near future.   
\end{abstract}

\section{Introduction}
Different experimental facilities like LHCb, COMPASS, BESIII, STAR, CMS, ATLAS, BABAR, etc., are continuously trying to produce enormous data in the field of heavy-flavor hadrons. The theoretical studies are also heavily invested in the field to interpret the experimental data and predict the properties of hadrons. The heavy quark hadrons like charmonium and bottomonium have also indeed received remarkable experimental growth. Through the efforts of various experimental collaborations and facilities, significant progress has been made in establishing bottomonium spectra. In this paper, we are focussing on the spectra and properties of bottomonium. The states $\Upsilon(1S)$, $\Upsilon(2S)$ and $\Upsilon(3S)$ were observed in 1977 by the E288 collaboration at the Fermi National Accelerator Laboratory (FNAL) \cite{herb1977,Innes1977} and notated as $\Upsilon, \Upsilon^{'}, \Upsilon^{''}$ states respectively. After the early detection of systems composing of $b\bar{b}$, many of the properties and new states were discovered as given in PDG\cite{PDG2022}. During the year 1982, the first excited state $\chi_{bJ}(2P)$ were identified in E1 transition from $\Upsilon^{''}$ state \cite{han1982,eigen1982}. Subsequently, in 1983, an exciting breakthrough came with the discovery of $\chi_{bJ}(2P)$ with J = 0, 1, 2 in the E1 transition from $\Upsilon^{'}$ state \cite{klopfenstein1983,fpauss1983}. In the following year, 1984, $\Upsilon(4S), \Upsilon(5S), \Upsilon(6S)$ are announced in $e^{+}e^{-}$ cross section above $B \bar B$ threshold \cite{besson1985,lovelock1985}. These findings opened the search window to investigate higher excited states of bottomonia spectra. After several experimental attempts, in 2008, the BABAR collaboration \cite{aubert2008} announced the observation of $\eta_{b}(1S)$, spin partner of $\Upsilon(1S)$. In 2010, BABAR collaboration observed $\Upsilon(1^{3} D_{J})$ with $J = 2$ in decay chain $\Upsilon(3S) \rightarrow \gamma \gamma \Upsilon(1^{3} D_{J}) \rightarrow \gamma \gamma \Upsilon(1S)$ \cite{del2010}. In 2011, the BABAR collaboration announced $h_{b}(1P)$ observed via decay process $\Upsilon(3S) \rightarrow \pi^{0}h_{b}(1P)$ \cite{lees2011}. In 2011, Belle Collaboration reported observation of states $h_{b}(1P)$ and $h_{b}(2P)$ in $e^{+}e^{-} \rightarrow h_{b}(nP)\pi^{+}\pi^{-}$ process \cite{adachi2012}. In the same year, the ATLAS group reconstructed $\chi_{b}(nP)$ via radiative decay $\chi_{b}(nP) \rightarrow \Upsilon(1S) \gamma$ and $\chi_{b}(nP) \rightarrow \Upsilon(2S) \gamma$. A new state $\chi_{b}(3P)$ is identified in these decay modes \cite{aad2012}. This state $\chi_{b}(3P)$ was also reconfirmed by D0 collaborations \cite{abazov2012} and the LHCb group \cite{aaij2014}. The latest CMS experiment shows resolved peaks of $\chi_{b1}(3P)$ and $\chi_{b2}(3P)$ with mass difference of $10.60\pm0.64(stat)\pm0.17(syst)$  $MeV$ using 80 $fb^{-1}$ in pp collisions at a center-of-mass energy of 13 $TeV$ \cite{sirunyan2018}. In 2012, Belle announced the first evidence of the existence of state $\eta_{b}(2S)$, pseudoscalar partner of $\Upsilon(2S)$ \cite{mizuk2012}. Some charged candidates like $Z_{b}(10610)$ and $Z_{b}(10650)$ are also announced in $\pi^{\pm} \Upsilon(nS)$(n = 1,2,3) and $\pi^{\pm} h_{b}(mP)$ ($m = 1, 2$) mass spectra by Belle experiment \cite{PDG2022}. But, their nature is still enigmatic. These states have been thoroughly investigated and analyzed in theoretical frameworks and interpreted these states as molecular states, exotic states, and tetraquarks states  \cite{wu2020, wang2014, chen2014}. Recently, there has been a resurgence of interest in the exploration and investigation of states $\Upsilon(10860)$ and $\Upsilon(11020)$, which was first announced in 1985 by CUSB experiment \cite{lovelock1985}. A recent study in 2019 by Belle investigated $\Upsilon(10860)$ state in process $e^{+}e^{-} \rightarrow \Upsilon(nS)$ ($n = 1, 2, 3$) with high significance and measured masses are  $10752.7\pm 5.9$ $MeV$,and decay widths are $35.5^{+17.6}_{-17.6}$ $MeV$ respectively \cite{mizuk2019}. Now this vector state is mentioned as $\Upsilon(10753)$ state in PDG. Authors in Ref.\cite{chen2020}, investigated $\Upsilon(10753)$ using relativistic flux tube and give assignment of $3^3 D_{1}$ $b \bar b$ state. In Ref.\cite{wang2019}, with QCD sum rules, Wang et al. interpreted this state as a hidden bottom tetraquark state. Theoretically, the prevailing consensus posits that $\Upsilon(10860)$  and $\Upsilon(11020)$ states correspond to the S-wave vector $b \bar b$ states denoted as $\Upsilon(5S)$ and $\Upsilon(6S)$, respectively \cite{deng2017,wang2018}. Authors in Ref.\cite{pandya2021,chaturvedi2020} explicated these states ($\Upsilon(10860)$ and $\Upsilon(11020)$) with instanton-induced potential obtained from the instanton liquid model for QCD vacuum and elucidated $\Upsilon(10860)$ as an admixture of $5^3S_{1} - 6^3D_{1}$ states and $\Upsilon(11020)$ as an admixture of $5^3S_{1} - 5^3D_{1}$ states respectively. Also state $\Upsilon(11020)$ studied in
Ref.\cite{pandya2021,chaturvedi2020} and suggested as an admixture of $6^3S_{1} - 5^3D_{1}$ states. Very recently, in March 2023, the BelleII detector examined the decay process $e^{+}e^{-} \rightarrow \omega_{\chi_{bJ}}(1P)$ ($J = 0, 1, 2$) and concluded states $\Upsilon(10860)$ and $\Upsilon(10753)$ may have different internal structures \cite{adachi2023}, which is considered same in previous studies \cite{mizuk2019}. All these recent observations of numerous states of heavy mesons developed an interest in theoreticians to explore them in all aspects. Different QCD potential models have studied all the above-observed states. It is believed that QCD potential models are the most successful phenomenological approach in the non-perturbative region. There are various kinds of potentials in literature like Cornell potential \cite{eichten1980,nrsoni2018,kher2022}, Martin potential \cite{martin1980,jena1983}, Logarithmic potential \cite{quigg1977}, Richardson potential \cite{richardson1979}, and Song and Lin potential \cite{song1987}, which successfully explore quarkonium and their properties. Another potential is used to study charmonium spectroscopy \cite{smruti2016} and its decay properties. In the same context, we explore bottomonium spectra and their decay properties using the same non-relativistic potential model from Ref. \cite{smruti2016}. Similar potentials have been harnessed with $n =2$ to study the dynamics of light hadrons in the framework of the Bethe- Salpeter equation under Covariant Instantaneous Ansatz (CIA) \cite{mittal1986, gupta1990, sharma1994}. It can extensively explore many processes like digamma decays, radiative decays, decay constants, and the structure of hadrons. This framework can also be explored for various energy scales to have better insights. Using quark - antiquark confining potential form \cite{smruti2016}, we calculated masses and decay properties of bottomonia - leptonic decays, gamma decays, gluons decays, and decay constants, which are very useful in revealing the non-perturbative domain of QCD.

The paper is summarized as follows: Section 2 gives a brief description of phenomenological quark-antiquark confining potential and extraction of potential parameters from Chi-square fitting. Section 3 gives a description of the decay rates of quarkonium. Section 4 presents the numerical analysis where we predict the mass spectra of bottomonia for $nS, nP, nD, nF, n = 1, 2, 3, 4, 5$  and their decay properties. Section 5 gives the conclusions of the paper.

\section{Framework}
Various theoretical approaches describe the spectrum of quarkonium states (charmonium, bottomonium, beauty charmed mesons). Still, the phenomenological potential model is one of the most popular and reliable approaches. The famous form of potential model exploited in heavy quarkonium spectroscopy is coulomb plus linear potential. The Coulomb potential component arises from the one-gluon exchange (Lorentz vector exchange) interaction between the quark and antiquark, which is analogous to the electromagnetic interaction between charged particles. On the other hand, the linear potential arises from the strong force (usually associated with Lorentz scalar exchange) that acts between the quark and the antiquark. This potential increases linearly with the separation between the quark and the antiquark. As discussed in the previous section, many other forms of confining potentials are used in literature. We have adopted the following potential form to study the spectroscopy of bottomonium bound states \cite{smruti2016}. 
\begin{align}
\label{pot}
    V(r) = V_{V} + V_{S} = \frac{-4 \alpha_{s}}{3r}+ \frac{A r^{2}}{(1+4B r^{p})^{\frac{1}{2}}}- V_{0}
\end{align}
where $V_{V}$ is the vector part of the potential (Coulomb), $V_{S}$ is the scalar part of the potential (confining), $\alpha_{s}$ is the running coupling constant. $A$ and $B$ are potential parameters estimated with chi-square fitting of low-lying bottomonium states. Since for charmonium, as mentioned in Ref.\cite{smruti2016}, the results are not consistent for $p=2$, we have also worked for $p =1$ and $B=1$ $GeV^p$. The value of the running coupling constant can be obtained by the expression :
\begin{align}
    \alpha_{s}(\mu^2) = \frac{4\pi}{\left(11- \frac{2}{3}n_{f}\right)\left(ln\frac{\mu^2}{\Lambda^2}\right)}
\end{align}
Where $\Lambda$ is the QCD scale taken as 0.12 $GeV$, $n_{f}$ is the number of active flavors, number of flavors lighter than scale $\mu$ ($m_{q}\ll\mu$). $\mu$ is the renormalization scale equal to $\frac{2m_{Q}m_{\bar Q}}{m_{Q}+ m_{\bar Q}}$, for bottomonium $n_{f} = 4$, as flavors $u$, $d$, $s$, and $c$ are considered light. 
%In this work, we have taken $\mu$ = 4.72 $GeV$%.
A similar potential is used to study the ground states of light-flavoured and heavy-light mesons \cite{Nsingh1988,CHAKRABARTY1989,anumittal1986,gupta1990,sharma1994,ALKOFER2001,alkofer2002,CVETIC2004,Bhatnagar_2006,bhatnagar2014}. The confining term in the potential Eq. \eqref{pot} is supposed to behave as linear confinement ($r$) for the heavy quark ($c$, $b$) and hadronic form of the ($r^2$) for the light quarks. %To study one gluon exchange interaction, $\alpha_{s}$ should be less than one, as for larger $\alpha_{s}$, perturbation theory will no longer be applicable. 
$V_{0}$ is a state-dependent constant potential. In adopted potential in eq \ref{pot}, we added spin-dependent terms - spin-spin, spin-orbit, and tensor to describe the spectrum's splitting structure and account for each state's different quantum numbers. The spin-dependent part $V_{SD}$ \cite{voloshin2008, berestetskii1982} is given by:
\begin{align}
    V_{SD} = V_{SS}\left[S(S+1)-\frac{3}{2}\right] &+ V_{LS}\left [\frac{1}{2}(J(J+1)-S(S+1)-L(L+1))\right]  \nonumber\\ 
    &+ V_{T}\left[S(S+1)-\frac{3(S_{1}.r)(S_{2}.r)}{r^{2}}\right] 
\end{align}
Where coefficients $V_{SS}$, $V_{LS}$, and $V_{T}$ depend on derivatives of vector $V_{V}$ and scalar $V_{S}$ contributions from the adopted potential in eq \ref{pot}. The expressions for coefficients are the following \cite{voloshin2008,gershtein1995}:
\begin{align}
     V_{SS}^{ij}(r)& = \frac{1}{3 M_{i} M_{j}} \nabla^{2}V_{V} = \frac{16 \pi \alpha_{s}}{9  M_{i} M_{j}} \delta^{3}(r)\\
      V_{LS}^{ij}(r) &= \frac{1}{2 M_{i} M_{j}r}\left[3\frac{dV_{V}}{dr}-\frac{dV_{S}}{dr}\right]\\
    V_{T}^{ij}(r) &= \frac{1}{6 M_{i} M_{j}}\left[\frac{d^{2}V_{V}}{dr^{2}}-\frac{1}{r}\frac{dV_{V}}{dr}\right]
    \end{align}
Where $M_{i}$, $M_{j}$ are quark masses. All these spin-dependent interactions are corrections in total mass, treated as first-order perturbation corrections in heavy quark-bound states. As $V_{LS}$, $V_{SS}$, and $V_{T}$ expressions are proportional to $1/m^{2}$ ($M_{i} = M_{j}= m$ for bottomonium), they justify their treatment as first-order perturbative corrections. The spin-orbit term containing $V_{LS}$ and tensor term containing $V_{T}$ describe the fine structure of the state, whereas the spin-spin term containing $V_{SS}$ describes hyperfine splittings. 
Mass is the prime property to study the spectroscopy of hadrons. To calculate spectra of bottomonium, we estimated the parameters $V_{0}$, bottom quark $m_{b}$, and parameter $A$ appearing in potential in eq \eqref{pot} with Chi-square fitting procedure. We considered these parameters as free parameters in the fitting procedure for ground-state bottomonium meson in the following range:
\begin{center}
        $0 < A < 1$ $(GeV^3)$; $0<V_{0} < 0.5$ $(GeV)$; $4 < m_{b}< 5$ $(GeV)$
\end{center}
The results for parameters are:
$m_{b}$ = 4.72 $GeV$, $V_{0}$ = 0.10 $GeV$, $A$ = 0.20 $GeV^{2}$. The running coupling constant is $\alpha_{s} = 0.2053$.
  %In this work, the mass of the bottom quark taken is 4.72 $GeV$%. 
The state dependent potential strength $V_{0}$ is given by relation:
\begin{align}
\label{ pot str}
    V_{0}(n+1, l) = V_{0}+b(n-1) +cl
\end{align}
Where $n$ and $l$ are the radial and angular quantum numbers of the states. Also, $b$ and $c$ are unknown parameters estimated by fitting the experimental masses of $2S$, $1P$ states of bottomonium. We find
$b$ = -0.105 $GeV$, $c$ = 0.0007 $GeV$. Using all these parameters and adopted potential form, we computed mass spectra of $nS$, $nP$, $nD$, and $nF$ bottomonium states listed in Table \ref{nS}, Table \ref{nP}, Table \ref{nD}, and Table \ref{nF}.

\section{Decay rates of heavy quarkonia}
Apart from the masses of bottomonium meson states, precise predictions of decay rates are crucial properties of any successful model. Recently there are several investigations on various phenomena encompassing strong, radiative, and leptonic decays of heavy quarkonium. Such investigations directly probe the hadron structure and shed light on some aspects of the quark-gluon structure. Leptonic decay constants serve as expedient probes for the short-distance structure of hadron and act as tools for studying quark dynamics in this domain. The extracted model parameters and radial wave functions are used here to compute the di-leptonic, two-photon, and two-gluon annihilation rates. Since these rates are related to the wave- function, they provide a better insight into quark-antiquark dynamics within mesons.
\subsection{Leptonic decay constants}
The study of leptonic decay constants of heavy quarkonia is crucial properties for understanding weak decays. It also provides information about CKM (Cabibbo-Kabayashi-Maskawa) matrix elements.  The expression for leptonic decay constants of pseudoscalar and vector mesons is given by \cite{van1967}:
\begin{align}
\label{decayconstant}
f^{2}_{P/V} = \frac{3|R_{ns P/V(0)}|^{2}}{\pi M_{ns P/V}} \bar C^{2}(\alpha_{s})
\end{align}
Here the QCD correction factor $\bar C^{2}(\alpha_{s})$ \cite{braaten1995,berezhnoy1996} is given by:
\begin{align}
\label{decayconstant1}
 \bar C^{2}(\alpha_{s}) = 1 - \frac{\alpha_{s}}{\pi} \left(\delta_{P,V} - \frac{m_{Q}- m_{\bar Q}}{m_{Q}+m_{\bar Q}}  ln \frac{m_{Q}}{m_{\bar Q}}\right)  
\end{align}
with $\delta_{P} = 2$ and $\delta_{V}= 8/3$ in case of Bottomonium mesons. The second term in equation \ref{decayconstant1} will disappear in the case of bottomonium.  Using these expressions, we computed leptonic decay constants and listed them in Table \ref{fp} and Table \ref{fv}.

\subsection{Electromagnetic Transition widths}
The study of electromagnetic transition can be understood in terms of electric and magnetic multipole expansion, and their investigation provides better sight into the non-perturbative regime. The selection rule for $E1$ transitions is $\Delta L = 0$ and $\Delta S = \pm 1$. For $M1$ transitions, $\Delta L = \pm1$ , and $\Delta S = 0$ are selection rules. To check the authenticity of calculated masses and chi-square fit parameters, we computed $E1$ and $M1$ transitions, and the formulas for these transitions \cite{eichten1975,eichten1978,brambilla2006,li2011} are given below:
\begin{align}
    \Gamma (n^{2S+1}L_{i}J_{i}\rightarrow n'^{2S+1}L_{f}J_{f}+ \gamma) &= \frac{4 \alpha_{e}\left<e_{Q}\right>^{2}\omega^{3}}{3}(2 J_{f}+1)S_{if}^{E1}|M_{if}^{E1}|^{2}\\
  \Gamma (n^{3}S_{1}\rightarrow n'^{1}S_{0}+ \gamma)& = \frac{ \alpha_{e}\mu^{2}\omega^{3}}{3}(2 J_{f}+1)|M_{if}^{M1}|^{2}  
    \end{align}
where, $\left<e_{Q}\right>$ is mean charge of $Q \bar Q$ system, $\mu$ is magnetic dipole moment and $\omega$ is photon energy and expression for them are following :
\begin{align}
    \left<e_{Q}\right> = \left|\frac{m_{\bar{Q} } e_{Q}-e_{\bar{Q} }m_{Q}}{m_{Q}+ m_{\bar{Q} }}\right|
\end{align}
\begin{align}
    \mu = \frac{e_{Q}}{m_{Q}} - \frac{e_{\bar{Q} }}{m_{\bar{Q} }}
\end{align}
and
\begin{align}
    \omega  = \frac{M_{i}^{2}- M_{f}^{2} }{2 M_{i}}
\end{align}
The symmetrical factor $S_{if}^{E1}$ is also given in the following way:
\begin{align}
    S_{if}^{E1} = max(L_{i},L_{f})\begin{pmatrix}
        J_{i}& 1& J_{f}\\
        L_{f}&S& L_{i}
    \end{pmatrix}^2
\end{align}
The matrix element $|M_{if}|$ for E1 and M1 transition can be written as 
\begin{align}
    |M_{if}^{E1}| = \frac{3}{\omega}\left<f\left|\frac{\omega r}{2}j_{0}\left(\frac{\omega r}{2}\right)-j_{1}\left(\frac{\omega r}{2}\right)\right|i\right>
\end{align}
and
\begin{align}
    |M_{if}^{M1}| = \left<f\left|j_{0}\left(\frac{\omega r}{2}\right)\right|i\right> 
\end{align}
The computed values of $E1$ and $M1$ transitions are collected in Table \ref{E1} and Table \ref{M1}, respectively.
\subsection{Leptonic decay widths}
Single virtual photons ($Q \bar Q \rightarrow l^{+}l^{-}$) are used to disintegrate quarkonium into leptons ($ e^{-}\mu^{-},\tau^{-}$). If the quarkonium state has the same quantum number as the photon, it decays to the lepton pair, i.e., $J^{PC} = 1^{--}$.  Using Van Royen Weisskopf formula, the leptonic decay width of $n^3S_{1}$ and $n^3D_{1}$ states of bottomonium, including first-order radiative QCD corrections \cite{van1967,kwong1988}, is given below:
\begin{align}
\label{leptoniceq1}
   \Gamma(n^3S_{1} \rightarrow e^{+}e^{-}) = \frac{4e_{Q}^{4} \alpha^{2} |R_{nS}(0)|^{2}}{M_{nS}^2}\left(1- \frac{16\alpha_{s}}{3\pi}\right)\\
   \label{leptoniceq2}
 \Gamma(n^3D_{1} \rightarrow e^{+}e^{-}) = \frac{25e_{Q}^{2} \alpha^{2} |R_{nD}^{''}(0)|^{2}}{2m_{Q}^{4}M_{nD}^2}\left(1- \frac{16\alpha_{s}}{3\pi}\right) 
\end{align}
where $e_{Q}$ is charge on quark, $M_{nS}$, $M_{nD}$ is masses of 
 decaying corresponding states, $\alpha = 1/137$ is fine structure constant that describes the strength of electromagnetic force, and $R_{nS}(0)$, $R_{nD}^{''}(0)$ are normalized reduced wave-function at origin for $S$ waves and second order derivative of normalized reduced wave function for $D$ waves respectively. Bracketed terms in expressions \ref{leptoniceq1}, \ref{leptoniceq2} are the lowest order QCD corrections. The computed values of leptonic decay widths are mentioned in Table \ref{Dilepton}.

\subsection{Digamma and trigamma decay widths}
A decay into two photons is forbidden to $J =1$ states by the Yang theorem \cite{landu1948, yang1950}. For other resonances, the conversation of charge parity requires the $S$ wave states to be in a spin-singlet state, whereas $P$
wave states of being in the spin-triplet state. The formula for digamma decay widths of $^1 S_{0}$ and $^3 P_{J} (J=0,2)$ is given below \cite{van1967, kwong1988}:
\begin{align}
\Gamma(n^1 S_0 \rightarrow \gamma \gamma) &= \frac{3 e^{2}_{Q}\alpha^{2}|R_{nS}(0)|^{2}}{m_{Q}^{2}}\left(1-\frac{3.4 \alpha_{s}}{\pi}\right)\\
\Gamma(n^3P_0 \rightarrow \gamma \gamma) &= \frac{27 e^{4}_{Q} \alpha^{2} |R^{'}_{nP}(0)|^{2} M_{^3P_{0}}}{2m_{Q}^{5}}\left[1+\left(\frac{\pi^{2}}{3}-\frac{28}{9}\right)\left(\frac{\alpha_{s}}{\pi}\right)\right]\\
\Gamma(n^3P_2 \rightarrow \gamma \gamma) &= \frac{4}{15}\frac{27 e^{4}_{Q} \alpha^{2} |R^{'}_{nP}(0)|^{2} M_{^3P_{2}}}{2m_{Q}^{5}}\left[1+\left(\frac{-16}{3}\right)\left(\frac{\alpha_{s}}{\pi}\right)\right]\\
\Gamma(n^3 S_1 \rightarrow \gamma \gamma \gamma) &= \frac{4(\pi^{2}-9) e^{6}_{Q}\alpha^{3}|R_{nS}(0)|^{2}}{m_{Q}^{2}}\left(1-\frac{12.6 \alpha_{s}}{\pi}\right)
\end{align}
$R^{'}_{nl}$ is the first derivative of the normalized reduced wave function at the origin in $P$ wave formulas. $e_{Q}$ is charge on quark. $\alpha = 1/137$ is a fine structure constant. Terms in brackets are the next to leading-order QCD radiative corrections. The calculated digamma, tri-gamma decay widths are collected in Table \ref{Digamma} and Table \ref{Trigamma}.
\subsection{Digluon decay widths}
At short range, two gluon decay widths are sensitive for quarkonia and its derivatives near the origin. The digluon decay width gives information about the total decay width of hadronic decay below the $Q \bar Q$ threshold. Employing VRW (Van Royen Weisskopf) method \cite{van1967, kwong1988} and including QCD radiative corrections \cite{lansberg2009,
barbieri1981,mangano1995}, expressions for digluon decay widths are:
\begin{align}
\Gamma(n^1S_0 \rightarrow gg) = \frac{2\alpha_s^2|R_{ns}(0)|^2}{3m_Q^2} \left(1 + \frac{4.4\alpha_s}{\pi}\right)
\end{align}
\begin{align}
     \Gamma(n^3P_{0}\rightarrow gg) &= \frac{6\alpha_{s}^{2}|R^{'}_{nP}(0)|^{2}}{m_{Q}^{4}}\left(1+ \frac{10.0 \alpha_{s}}{\pi} \right)
\end{align}
      
  \begin{align}
       \Gamma(n^3P_{2}\rightarrow gg) &= \frac{8\alpha_{s}^{2}|R^{'}_{nP}(0)|^{2}}{5m_{Q}^{4}}  \left(1- \frac{0.1 \alpha_{s}}{\pi} \right)\\
\Gamma(n^1D_{2}\rightarrow gg) &= \frac{2\alpha_{s}^{2}|R^{''}_{nD}(0)|^{2}}{3 \pi m_{Q}^{6}}
\end{align}
Terms in brackets are the next to leading-order QCD radiative corrections. $\alpha = 1/137$ is fine structure constant. $R^{'}_{nP}$ is the first order derivative of the normalized reduced wave function at the origin in $P$ wave formulas. Similar $R^{"}_{nD}$ is the second order derivative of the normalized reduced wave function at the origin in $D$ wave formulas. The calculated values of digluons are mentioned in Table \ref{Digluon}.

\section{Results and Discussions}
\subsection{Mass Spectroscopy}
The spectroscopic masses of $S, P, D, F$ waves of bottomonium mesons are calculated by employing a non-relativistic potential given in Eq. \eqref{pot} and solving the Schrodinger wave equation numerically. Our calculated mass for $S, P, D, F$ waves of bottomonium mesons is listed in Table \ref{nS}, Table \ref{nP}, Table \ref{nD}, Table \ref{nF} respectively and compared with experimental data as well as different theoretical predictions. 
\begin{table}[ht!] 
\centering
\caption{\label{nS} The predicted values of $nS$ bottomonium masses ($GeV$) compared with some other model predictions.}
\addtolength{\leftskip} {-2cm}
    \addtolength{\rightskip}{-2cm}
 \begin{tabular*}{1.2\linewidth}{@{\extracolsep{\fill}}ccccccccccc}\toprule
$J^{PC}(n^{2S+1}L_{j})$ & Ours &Ref.\cite{kher2022} & Ref.\cite{wang2018} &Ref.\cite{godfrey2015}& Ref.\cite{pandya2021}&Ref.\cite{shah2012}& Ref. \cite{ebert2011}&Ref.\cite{radford2011}& PDG \cite{PDG2022}\\ \midrule
$0^{-+}(1^{1}S_{0})$ & 9.43297 & 9.423 & 9.398 & 9.402 & 9.41222 & 9.392 & 9.398 & 9.393 & 9.399\\
$1^{--}(1^{3}S_{1}) $ & 9.44816 & 9.463 & 9.463 & 9.465 & 9.46075 & 9.460 & 9.460 & 9.460 & 9.460  \\
$0^{-+}(2^{1}S_{0})$ & 10.00660 & 9.983 & 9.989 & 9.976 & 9.99548 & 9.991 & 9.990 & 9.987 & 9.999 \\
$1^{--}(2^{3}S_{1})$ & 10.02020 & 10.001 & 10.017 & 10.003 & 10.02622 & 10.024 & 10.023 & 10.023 & 10.023 \\
$0^{-+}(3^{1}S_{0})$ & 10.50310 & 10.342 & 10.336 & 10.336 & 10.33900 & 10.323 & 10.329 & 10.345& - \\
$1^{--}(3^{3}S_{1})$ & 10.51650 & 10.354 & 10.356 & 10.354 & 10.36465 & 10.346 & 10.355 & 10.364 & 10.355\\
$ 0^{-+}(4^{1}S_{0})$ & 10.96910 & 10.638 & 10.597 & 10.623 & 10.57249 & 10.558 & 10.573 & 10.364 & - \\
$1^{--}(4^{3}S_{1})$ & 10.98250 & 10.650 & 10.612 & 10.635 & 10.59447 & 10.575 & 10.586 & 10.643 & 10.579 \\
$ 0^{-+}(5^{1}S_{0})$ & 11.41690 & 10.901 & 10.810 & 10.869 & 10.74676 & 10.741 & 10.851 & - & - \\
$1^{--}(5^{3}S_{1}) $ & 11.43030 & 10.912 & 10.822 & 10.878 & 10.76614 & 10.755 & 10.869 & - & 10.885 \\ \bottomrule
\end{tabular*}
\end{table}
%%%%%%%%%%%%%%%%%%%%%%%%%%%%%%%%%%%%%%%%%%%%%%%%%%%%%%%%%%%%%%%%%%%%%%%%%%%%%%%%%%%%%%%
\begin{table}[ht!]

\caption{ \label{nP}The predicted values of bottomonium $nP$ meson masses ($GeV$) compared with some other model predictions.}
\addtolength{\leftskip} {-2cm}
    \addtolength{\rightskip}{-2cm}
 \begin{tabular*}{1.2\linewidth}{@{\extracolsep{\fill}}ccccccccccc}\toprule
$J^{PC}(n^{2S+1}L_{j})$ & Ours
  &Ref.\cite{wang2018}
 & Ref.\cite{godfrey1985} & Ref.\cite{pandya2021}&Ref. \cite{nrsoni2018}&Ref. \cite{godfrey2015}&Ref.\cite{ebert2011}&Ref. \cite{li2009}& PDG\cite{PDG2022}\\
 \hline
$ 0^{++}(1^{3}P_{0})$&9.75011&9.858&9.845&9.84961&9.806&9.847&9.859&9.865& 9.859 \\
$ 1^{+-}(1^{1}P_{1})$ &9.75938 & 9.894& 9.881 & 9.87456&9.819&9.876&9.892&9.897&9.899 \\
$ 1^{++}(1^{3}P_{1})$ &9.75671&9.889&9.875&9.87147&9.821&9.882&9.900&9.903&9.893  \\
$2^{++}(1^{3}P_{2})$&9.76608&9.910&9.896&9.88140&9.825&9.897&9.912&9.918& 9.912 \\
$ 0^{++}(2^{3}P_{0})$&10.25270&10.235&10.225&10.25254&10.205&10.226&10.233&10.226&10.233 \\
$ 1^{+-}(2^{1}P_{1})$ &10.26000&10.259&10.250&10.27000&10.217&10.246&10.255&10.251& 10.260 \\
$ 1^{++}(2^{3}P_{1})$ &10.25840&10.255&10.246&10.26786&10.220&10.250&10.260&10.256&10.255  \\
$2^{++}(2^{3}P_{2})$&10.26670&10.269&10.261&10.27477&10.224&10.261&10.268&10.269& 10.269 \\

$ 0^{++}(3^{3}P_{0})$&10.72320&10.513&10.521&10.51288&10.540&10.552&10.521&10.502&-\\
$ 1^{+-}(3^{1}P_{1})$ &10.72930&10.530&10.540&10.52650&10.553&10.538&10.541&10.524&-\\
$ 1^{++}(3^{3}P_{1})$ &10.72840&10.527&10.537&10.52484&10.556&10.541&10.544&10.529&10.514 \\
$2^{+-}(3^{3}P_{2})$&10.73630&10.539&10.549&10.53021&10.560&10.550&10.550&10.540&10.524 \\

$ 0^{++}(4^{3}P_{0})$&11.17450&10.736&10.773&10.70356&10.840&10.775&10.781&10.732&-\\
$ 1^{+-}(4^{1}P_{1})$ &11.17990&10.751&10.790&10.71480&10.853&10.788&10.802&10.753&-\\
$ 1^{++}(4^{3}P_{1})$ &11.17950&10.749&10.787&10.71344&10.855&10.790&10.804&10.757&-\\
$2^{++}(4^{3}P_{2})$&11.18700&10.758&10.797&10.71786&10.860&10.798&10.812&10.767&-\\

$ 0^{++}(5^{3}P_{0})$&11.61260&10.926&10.998&10.85338&11.115&11.014&-&10.933&-\\
$ 1^{+-}(5^{1}P_{1})$ &11.61730&10.938&11.013&10.86300&11.127&11.014&-&10.951&-\\
$ 1^{++}(5^{3}P_{1})$ &11.61740&10.936&11.010& 10.86183&11.130&11.016&-&10.955&-\\
$2^{++}(5^{3}P_{2})$&11.62460&10.944&11.020&10.86562&11.135&11.022&-&10.965&-\\

\hline
\end{tabular*}
\end{table}
%%%%%%%%%%%%%%%%%%%%%%%%%%%%%%%%%%%%%%%%%%%%%%%%%%%%%%%%%%%%%%
%%%%%%%%%%%%%%%%%%%%%%%%%%%% nD %%%%%%%%%%%%%%%%%%%%%%%%%%%%%%
%%%%%%%%%%%%%%%%%%%%%%%%%%%%%%%%%%%%%%%%%%%%%%%%%%%%%%%%%%%%%%
\begin{table}[ht!]
\caption{\label{nD}The predicted values of bottomonium $nD$ meson masses ($GeV$) compared with some other model predictions.}
\addtolength{\leftskip} {-2cm}
    \addtolength{\rightskip}{-2cm}
 \begin{tabular*}{1.2\linewidth}{@{\extracolsep{\fill}}ccccccccccc}\toprule
$J^{PC}(n^{2S+1}L_{j})$ & Ours
  &Ref.\cite{wang2018}
 & Ref.\cite{godfrey1985} & Ref. \cite{pandya2021}&Ref.\cite{shah2012}&Ref. \cite{li2009}&Ref. \cite{ebert2011}&Ref.\cite{nrsoni2018}& PDG \cite{PDG2022}\\
 \midrule
$ 1^{--}(1^{3}D_{1})$&9.99397&10.153&10.137&10.14499&10.147&10.145&10.154&10.074&-\\
$ 2^{-+}(1^{1}D_{2})$&9.99238&10.163&10.148&10.15380&10.162&10.151&10.161&10.075&-\\
$ 2^{--}(1^{3}D_{2})$ &9.99423&10.162&10.147&10.15277&10.166&10.152&10.163&10.074& 10.164 \\
$ 3^{--}(1^{3}D_{3})$ &9.99386&10.170&10.155&10.15831&10.177&10.156&10.166&10.073&-\\

$ 1^{--}(2^{3}D_{1})$&10.46920&10.442&10.441&10.45023&10.428&10.432&10.435&10.423&-\\
$ 2^{-+}(2^{1}D_{2})$&10.46740&10.450&10.450&10.45660&10.437&10.438&10.443&10.424&-\\
$ 2^{--}(2^{3}D_{2})$ &10.46990&10.450&10.449&10.45586&10.440&10.439&10.445&10.424&-\\
$ 3^{--}(2^{3}D_{3})$ &10.47040&10.456&10.455&10.45985&10.447&10.442&10.449&10.423&-\\

$ 1^{--}(3^{3}D_{1})$&10.92470&10.675&10.698&10.65968&10.637&10.670&10.704&10.731&-\\
$ 2^{-+}(3^{1}D_{2})$&10.92210&10.681&10.706&10.66470&10.645&10.676&10.711&10.733&-\\
$ 2^{--}(3^{3}D_{2})$ &10.92550&10.681&10.705&10.66412&10.646&10.677&10.713&10.733&-\\
$ 3^{--}(3^{3}D_{3})$ &10.92640&10.686&10.711&10.66725&10.652&10.680&10.717&10.733&-\\

$ 1^{--}(4^{3}D_{1})$&11.36630&10.871&10.927&10.81883&10.805&10.877&10.949&11.013&-\\
$ 2^{-+}(4^{1}D_{2})$&11.36290&10.876&10.934&10.82300&10.811&10.882&10.957&11.016&-\\
$ 2^{--}(4^{3}D_{2})$ &11.36710&10.876&10.934&10.82252&10.813&10.883&10.959&11.015&-\\
$ 3^{--}(4^{3}D_{3})$ &11.36810&10.880&10.939&10.82512&10.817&10.886&10.963&11.015&-\\

$ 1^{--}(5^{3}D_{1})$&11.79730&11.041&11.137&10.94901&10.945&11.060&-&-&-\\
$ 2^{-+}(5^{1}D_{2})$&11.79300&11.046&11.143&10.95260&10.952&11.066&-&-&-\\
$ 2^{--}(5^{3}D_{2})$ &11.79820&11.045&11.143&10.95159&10.950&11.065&-&-&-\\
$ 3^{--}(5^{3}D_{3})$ &11.79920&11.049&11.148&10.95442&10.955&11.069&-&-&-\\
\bottomrule
\end{tabular*}
\end{table}
%%%%%%%%%%%%%%%%%%%%%%%%%%%%%%%%%%%%%%%%%%%%%%%%%%%%%%%%%%%%% 
\begin{table}[ht!]
\centering
\caption{\label{nF} The predicted values of bottomonium $nF$ meson masses ($GeV$) compared with some other model predictions.}
 \addtolength{\leftskip} {-2cm}
    \addtolength{\rightskip}{-2cm}
 \begin{tabular*}{1.2\linewidth}{@{\extracolsep{\fill}}ccccccccccc}\toprule
$J^{PC}(n^{2S+1}L_{j})$ & Ours
  &Ref.\cite{wang2018}
 & Ref.\cite{godfrey1985} & Ref. \cite{godfrey2015}&Ref. \cite{nrsoni2018}& Ref. \cite{ebert2011}&Ref. \cite{radford2011}\\
 \midrule
$ 2^{++}(1^{3}F_{2})$&10.20690&10.362&10.350&10.358&10.283&10.343&10.353\\
$ 3^{+-}(1^{1}F_{3})$&10.20130&10.366&10.354&10.355&10.288&10.347&10.356\\
$ 3^{++}(1^{3}F_{3})$ &10.20430&10.366&10.354&10.355&10.287&10.346&10.356\\
$ 4^{++}(1^{3}F_{4})$ &10.20050&10.369&10.358&10.358&10.291&10.349&10.357\\

 $ 2^{++}(2^{3}F_{2})$&10.66600&10.605&10.615&10.615&10.604&10.610&10.610\\
$ 3^{+-}(2^{1}F_{3})$&10.66080&10.609&10.619&10.619&10.607&10.647&10.613\\
$ 3^{++}(2^{3}F_{3})$ &10.66500&10.609&10.619&10.619&10.607&10.614&10.613\\
$ 4^{++}(2^{3}F_{4})$ &10.66340&10.612&10.622&10.622&10.609&10.617&10.615\\
$ 2^{++}(3^{3}F_{2})$&11.11170&10.809&10.849&10.850&10.894&-&-\\
$ 3^{++}(3^{1}F_{3})$&11.10530&10.812&10.853&10.853&10.897&-&-\\
$ 3^{++}(3^{3}F_{3})$ &11.11120&10.812&10.853&10.853&10.896&-&-\\
$ 4^{++}(3^{3}F_{4})$ &11.11040&10.815&10.856&10.856&10.898&-&-\\

$ 2^{++}(4^{3}F_{2})$&11.54670&10.985&11.063&-&-&-&-\\
$ 3^{+-}(4^{1}F_{3})$&11.53850&10.988&11.066&-&-&-&-\\
$ 3^{++}(4^{3}F_{3})$ &11.54640&10.988&11.066&-&-&-&-\\
$ 4^{++}(4^{3}F_{4})$ &11.54590&10.990&11.066&-&-&-&-\\
\bottomrule
\end{tabular*}
\end{table}
 
Our findings for the masses concur with experimentally observed masses for respective states. Our calculated masses for $1^3S_{1}$ state $9.44816$ $GeV$, which is found to be in good agreement with experimental value $9.460 \pm 0.00026$ $GeV$ and for its spin partner $1^1S_{0}$ state, our estimated mass is roughly 11 $MeV$ higher than experimental mass mentioned in PDG. On observing Table \ref{nS}, we found masses of nS wave up to n = 5 agree with the experimental data and other theoretical predictions with a deviation of $\pm 5\%$. For $nS$ states, our predicted masses are close to theoretical estimations of Ref. \cite{kher2022, godfrey2015, shah2012}. In Ref. \cite{kher2022}, Kher et al. employed a variational method with a single Gaussian trial wave functions, Ref. \cite{godfrey2015}, Godfrey et al. employed relativized quark model and Ref. \cite{shah2012}, Shah et al. employed the non-relativistic potential model to investigate bottomonia mass spectroscopy. Specifically, on comparing with Ref. \cite{wang2018}, we observed our masses are very close to their predictions up to $n = 4$ $S$ wave states, but our results slightly deviate for $5 S$ states from them. This may happen due to screening effects employed by Ref. \cite{wang2018} as screening effects contribute significantly to higher excited states and affects mass values, wave functions, and decay behaviors. 
 Similarly, for $P$ waves, our results are in good agreement with experimental values and theoretical estimations. For $n = 1$ $P$ wave masses, our mass estimations are $1.45\%$ below experimental values. Our computation for state $2^{1}P_{1}$ is $10.2600$, which is in excellent agreement with the experimental value mentioned in PDG ($10.260 \pm 1.2\times 10^{-4}$). Also, its spin partners, i.e $2^{3}P_{0}, 2^{3}P_{1}, 2^{3}P_{2}$, our results differ from experimental values by $19$ $MeV$, $3$ $MeV$, $2$ $MeV$ respectively which reflects reliability of our model. One more important parameter to check the authenticity of any model is hyperfine splittings. For $1P$ state, $\Delta M _{hfs}(1P)$ = 4.9 $MeV$ and $2P$ state $\Delta M _{hfs}(2P)$ = 2.4 $MeV$, which consistent to experimental values \cite{PDG2022}. This splitting decreases more and becomes negligible for higher excited states. This shows that spin-spin contribution is negligible for higher excited states and thus reflects the vanishing of long-range chromomagnetic interactions in quarkonium. The calculated masses of $n = 4,5$ for $P$ are found to be overestimated in comparison to other theoretical estimations. However, masses for $n =5$ P wave are closer to outcomes of Ref. \cite{godfrey2015,godfrey1985,nrsoni2018} than other estimations. In Ref. \cite{godfrey1985,godfrey2015}, Godfrey et al. implemented a relativistic approach, while Ref. \cite{nrsoni2018} used non-relativistic approach. This shows relativistic factors have very small contributions that do not much affect the mass spectroscopy of bottomonia. In Table \ref{nD}, we listed predicted masses for $nD$ wave ($ n =1, 2, 3, 4, 5$) and compared them with other model predictions and experimental masses, which are also in good accord with them. Only one state $1^1 D_{2}$ of $D$ waves is experimentally available and the measured value of mass is $10.164$ $GeV$. Our predicted value for this state deviates by $1.59 \%$ from the experimental value. Our result for $2D$ masses is found to be consistent with other theoretical predictions, while masses of $1D, 3D, 4D $ states are higher than predictions of other models. However, masses for $5D$ states are closer to predictions of Ref. \cite{godfrey1985}, which employed relativized quark model with chromodynamics and included the concept of one-gluon-exchange-plus-linear confinement potential. This is also notable in Ref. \cite{godfrey1985}, authors have not taken spin-spin, spin-orbit interaction into account, which reflects in mass degeneracy in spin states. 
Also, computed masses for $nF$ wave ($n =1,2,3,4$) listed in Table \ref{nF} are very close to other models' predictions. For the $1F$ state, our results are more consistent with Ref. \cite{nrsoni2018}, which used Cornell's potential to study bottomonium spectra. This is a similar kind of method that we used in this paper. Such comparison enhances the validity of our framework. Other results of $F$ states are also in reasonable agreement with theoretical estimates. 
\subsection{Decay constants}
The study of leptonic decay constants is important for investigating non - perturbative QCD dynamics. Using calculated masses for S-wave collected in Table \ref{nS} and applying Van Royen-Weisskopf formula, extracted in eqns \ref{decayconstant},  \ref{decayconstant1}, we computed values of pseudoscalar decay constants $f_{P}$ and vector decay constants $f_{V}$ listed in Table \ref{fp} and Table \ref{fv} respectively. 
\begin{table}[ht!]
\centering
\caption{ \label{fp} Pseudoscalar decay constants of bottomonium ($MeV$) without and with QCD corrections}
\addtolength{\leftskip} {-2cm}
    \addtolength{\rightskip}{-2cm}
 \begin{tabular*}{1\linewidth}{@{\extracolsep{\fill}}ccccccccc}\toprule
 States& $f_{P}$ & $f_{P}^C$ 
  &Ref.\cite{kher2022}
 & Ref.\cite{pandya2021} & Ref.\cite{nrsoni2018}&Ref.\cite{krassnigg2016}&Ref. \cite{patel2009}&Ref. \cite{wang2006}\\
 \midrule
$ 1S$&632.626& 549.879 &529&578.21&646.025&756&744&498\\
 $2S $&551.298& 479.188 &317&499.48&518.803&285&577&366\\
$ 3S$&528.840& 459.668 &280&450.35&474.954&333&511&304\\
 $4S $&516.337& 448.800 &264&413.93&449.654&40&471&259\\
$ 5S$&507.351& 440.990 &255&385.68&432.072&-&443&228\\
 
 \bottomrule
\end{tabular*}
\end{table}

\begin{table}[ht!]
\centering
\caption{\label{fv} Vector decay constants of bottomonium ($MeV$) without and with QCD corrections}
 \addtolength{\leftskip} {-2cm}
    \addtolength{\rightskip}{-2cm}
 \begin{tabular*}{1.1\linewidth}{@{\extracolsep{\fill}}cccccccccc}\toprule
 States&$f_{V}$&$f_{V}^C$& Ref.\cite{kher2022}
  &Ref.\cite{pandya2021}
  & Ref.\cite{nrsoni2018}&Ref. \cite{wang2006}& Ref. \cite{bhavsar2018}& Ref. \cite{patel2009}& PDG \cite{PDG2022}\\
 \midrule
$ 1S$&666.036& 535.517 &530&551.53&647.250 &498&705.4&706& 715 \\
 $2S $&564.438& 466.707 &317&477.05&519.436&366&554.9&547& 498 \\
$ 3S$&542.290& 447.715 &280&430.42&475.440&304&436.8&484& 430\\
 $4S $&529.483& 437.141 &265&395.80&450.066&259&332.4&446& 336\\
$ 5S$&520.281& 429.544 &255&368.91&432.437&228&286.5&419&-\\
 
 \hline
\end{tabular*}
\end{table}
Our results for pseudoscalar decay constants are in tune with Ref. \cite{kher2022, pandya2021, nrsoni2018}, while lower from Ref. \cite{krassnigg2016, patel2009}. It is noticed that Ref. \cite{kher2022, pandya2021, nrsoni2018} have employed a similar kind of framework with different potential forms, which enhance the authenticity of our work. The estimated values of vector decay constant for states $1S, 2S, 3S, 4S $ differ by 179 $MeV$, 31 $MeV$, 17 $MeV$, and 101 $MeV$ from experimental values, respectively. In comparison with theoretical approaches, values for vector decay constants without QCD corrections are consistent with Ref. \cite{nrsoni2018}, while with QCD corrections, our values concur with Ref. \cite{pandya2021}. Ref. \cite{pandya2021}  employed instanton-induced potential by including confining terms to study bottomonium spectroscopy.
\subsection{Radiative Transitions}
The numerical result of radiative transitions in electric diploe (E1) and magnetic dipole (M1) expansion are listed in Table \ref{E1}and Table \ref{M1}, respectively.
\begin{table}[ht!]
\centering
\caption{\label{E1} E1 transition widths of bottomonia (in $keV$)}
\addtolength{\leftskip} {-2cm}
    \addtolength{\rightskip}{-2cm}
 \begin{tabular*}{1.1\linewidth}{@{\extracolsep{\fill}}cccccccc}\toprule
 Decays&Ours& Ref.\cite{nrsoni2018}
  & PDG\cite{PDG2022}
 & Ref.\cite{deng2017}&Ref. \cite{radford2007}& Ref. \cite{ebert2011}& Ref. \cite{li2009}  \\
 \midrule
$ 2^{3}S_{1}\rightarrow 1^{3}P_{0}$&1.38000&2.370&1.220 & 1.09&1.15&1.65&1.67\\
$ 2^{3}S_{1}\rightarrow 1^{3}P_{1}$&3.87511&5.689&2.210 &2.17&1.87&2.57&2.54\\
$ 2^{3}S_{1}\rightarrow 1^{3}P_{2}$&5.84652&8.486&2.290 &2.62&1.88&2.53&2.62\\
$ 2^{3}S_{0}\rightarrow 1^{1}P_{1}$&10.3487  &10.181&-&3.41&4.17&3.25&6.10\\
 $ 3^{3}S_{1}\rightarrow 2^{3}P_{0}$&1.44755&3.330&1.200 &1.21&1.67&1.65&1.83\\
$ 3^{3}S_{1}\rightarrow 2^{3}P_{1}$&4.10768&7.936&2.560 &2.61&2.74&2.65&2.96\\
$ 3^{3}S_{1}\rightarrow 2^{3}P_{2}$&6.29319&11.447&2.660&3.16&2.80&2.89&3.23\\
$ 3^{3}S_{1}\rightarrow 1^{3}P_{0}$ &6.89619&0.594&0.055 &0.097&0.03&0.124&0.07\\
$ 3^{3}S_{1}\rightarrow 1^{3}P_{1}$&2.54430&1.518&0.018&0.0005&0.09&0.307&0.17\\
$ 3^{3}S_{1}\rightarrow 1^{3}P_{2}$&3.91653&2.354&0.200 &0.14&0.13&0.445&0.15\\
$ 3^{1}S_{0}\rightarrow 1^{1}P_{1}$&6.96943&3.385&-&0.67&0.03&0.770&1.24\\
$ 3^{3}S_{0}\rightarrow 2^{1}P_{1}$&11.39010&13.981&-&4.25&-&3.07&11.0\\
$ 1^{3}P_{2}\rightarrow 1^{3}S_{1}$& 37.69970 &57.530&34.380&31.8&31.2&29.5&38.2\\
$ 1^{3}P_{1}\rightarrow 1^{3}S_{1}$& 34.72050 &54.927&32.544&31.9&27.3&37.1&33.6\\
$ 1^{3}P_{0}\rightarrow 1^{3}S_{1}$& 32.70690 &49.530&-&27.5&22.1&42.7&26.6\\
$ 1^{1}P_{1}\rightarrow 1^{1}S_{0}$& 38.71140 &72.094&35.770&35.8&37.9&54.4&55.8\\
$ 2^{3}P_{2}\rightarrow 2^{3}S_{1}$& 11.05290 &28.848&15.100 &15.5&16.8&18.8&18.8\\
$ 2^{3}P_{1}\rightarrow 2^{3}S_{1}$& 10.07620 &26.672&19.400&15.3&13.7&15.9&15.9\\
$ 2^{3}P_{0}\rightarrow 2^{3}S_{1}$ & 9.43571 &23.162&-&14.4&9.90&11.7&11.7\\ 
$ 2^{1}P_{1}\rightarrow 2^{1}S_{0}$ & 11.01620 &35.578&-&16.2&-&23.6&24.7\\ 
$ 2^{3}P_{2}\rightarrow 1^{3}S_{1}$ & 9.41195 &29.635&9.800 &12.5&7.74&8.41&13.0\\ 
$ 2^{3}P_{1}\rightarrow 1^{3}S_{1}$ & 9.20092 &28.552&8.900 &10.8&7.31&8.01&12.4\\ 
$ 2^{3}P_{0}\rightarrow 1^{3}S_{1}$ & 9.05706 &28.552&-&10.8&6.69&7.36&11.4\\ 
$ 2^{1}P_{1}\rightarrow 1^{1}S_{0}$ & 9.44253 &26.769&-&5.4&-&9.9&15.9\\
$ 1^{3}D_{1}\rightarrow 1^{3}P_{0}$ & 16.24530 &34.815&-&16.1&-&24.2&23.6\\ 
$ 1^{3}D_{1}\rightarrow 1^{3}P_{1}$ & 11.27790 &9.670&-&19.8&-&12.9&12.3\\ 
$ 1^{3}D_{1}\rightarrow 1^{3}P_{2}$ & 0.67091 & 0.394 &-& 13.3&-&0.67&0.65\\ 
$ 1^{3}D_{2}\rightarrow 1^{3}P_{1}$ & 20.36500 & 11.489 &-&1.02&19.3&24.8&23.8\\ 
$ 1^{3}D_{2}\rightarrow 1^{3}P_{2}$ & 6.05832 &3.583&-&7.23&5.07&6.45&6.29\\ 
$ 1^{3}D_{3}\rightarrow 1^{3}P_{2}$ & 24.12070 &14.013&-&32.1&21.7&26.7&26.4\\ 
$ 1^{1}D_{2}\rightarrow 1^{1}P_{1}$ & 25.03720 &14.821&-&30.3&-&30.2&42.3\\ 
\bottomrule
\end{tabular*}
\end{table}
%%%%%%%%%%%%%%%%%%%%%%%%%%%%%%%%%%%%%%%%%%%%%%%%%%%%%%%%%%%%%%%%%%%%%%%%%%%%%
 \begin{table}[ht!] 
 \centering
 \caption{\label{M1} M1 transition widths of bottomonia (in eV)}
 \addtolength{\leftskip} {-2cm}
    \addtolength{\rightskip}{-2cm}
 \begin{tabular*}{1.1\linewidth}{@{\extracolsep{\fill}}cccccccc}\toprule
Decays& Ours & Ref.\cite{kher2022}
  &Ref.\cite{deng2017}
 &PDG\cite{PDG2022}&Ref. \cite{radford2007}& Ref. \cite{ebert2003}& Ref. \cite{bhaghyesh2011} \\
 \hline
$ 1^{3}S_{1}\rightarrow 1^{1}S_{0}$&0.01436&37.668&10.00&-&4.00&5.8&15.36\\
$ 2^{3}S_{1}\rightarrow 2^{1}S_{0}$&0.01601&5.619&0.59&-&0.05&1.4&1.82\\
$ 2^{3}S_{1}\rightarrow 1^{1}S_{0}$&1.25374&77.173&66.00&$12.5 \pm 4.9$&-&6.4&-\\
 $ 3^{3}S_{1}\rightarrow 3^{1}S_{0}$&0.01629&2.849&3.90&-&-&0.8&-\\
$ 3^{3}S_{1}\rightarrow 2^{1}S_{0}$&3.12243&36.177&11.00& $< 14$&-&1.5&-\\
$ 3^{3}S_{1}\rightarrow 1^{1}S_{0}$&1.72572&76.990&71.00&$10 \pm 2$&-&10.5&-\\
\hline
\end{tabular*}
\end{table}
%%%%%%%%%%%%%%%%%%%%%%%%%%%%%%%%%%%%%%%%%%%%%%%%%%%%%%%%%%%%%%%%%%%%%%%%%%%%%%%%%%%%%%%%%%%%%%%%%%
Our estimations for radiative transitions widths $\Gamma( 2^{3}S_{1}\rightarrow 1^{3}P_{0}\gamma)$,
$\Gamma( 2^{3}S_{1}\rightarrow 1^{3}P_{1}\gamma) $,  $\Gamma (2^{3}S_{1}\rightarrow 1^{3}P_{2}\gamma)$ differ by 0.16 $keV$, 1.66 $keV$, 3.5 $keV$ from experimental values, respectively, and are also compatible with theoretical models estimations. These transitions ($\Gamma( 2^{3}S_{1}\rightarrow 1^{3}P_{0}\gamma)$, $\Gamma( 2^{3}S_{1}\rightarrow 1^{3}P_{1}\gamma) $, $\Gamma (2^{3}S_{1}\rightarrow 1^{3}P_{2}\gamma)$) are closer to results of the potential model with $v^{2}/c^{2}$ corrections \cite{radford2007}, quasi potential approach \cite{ebert2011}, the potential model with screened potential \cite{li2009} shown in Table \ref{E1}. The E1 transitions  $ \Gamma(3^{3}S_{1}\rightarrow 2^{3}P_{0}\gamma)$, $\Gamma(3^{3}S_{1}\rightarrow2^{3}P_{1}\gamma)$, $\Gamma(3^{3}S_{1}\rightarrow 2^{3}P_{2}\gamma)$ are consistent with experimental data while $ \Gamma(3^{3}S_{1}\rightarrow 1^{3}P_{0}\gamma)$, $ \Gamma(3^{3}S_{1}\rightarrow 1^{3}P_{1}\gamma)$, $ \Gamma(3^{3}S_{1}\rightarrow 1^{3}P_{2}\gamma)$ are lower by 7 $keV$, 3 $keV$, 4 $keV$ from experimental results respectively. It is also inferred from Table \ref{E1}, radiative transition widths for $3P \rightarrow 1S$ are suppressed than other $E1$ transitions. This shows a general feature that $E1$ transition between two states differs by two radial numbers that are highly suppressed and have very less contribution in $E1$ transitions. The calculated electric dipole transitions for $\Gamma (1^{3}P_{2}\rightarrow 1^{3}S_{1}\gamma)$, $ \Gamma(1^{3}P_{1}\rightarrow 1^{3}S_{1}\gamma)$, $ \Gamma(1^{1}P_{1}\rightarrow 1^{1}S_{0}\gamma)$ are very close to experimental data. Also, the $E1$ transitions for  $1P \rightarrow 1S$ give higher contributions relative to $2P \rightarrow 1S$ transition, which is  expected. For $E1$ transition  $1D \rightarrow 1P$, there is no experimental data, but compared with theoretical predictions, it is found that our results have a high degree of similarity with Ref. \cite{ebert2011, li2009}, while lower values than Ref. \cite{nrsoni2018}.
We also calculated $M1$ transitions and listed them in Table \ref{M1}. Generally, $M1$ transitions are weaker than $E1$ transitions but play an important role in finding spin-singlet states, which is back-breaking in other ways. It is inferred from Table \ref{M1} that predicted values are very small relative to $E1$ transitions. Our results for $M1$ transitions are lower than other model's predictions.

\subsection{Annihilation Decays}
The study of annihilation decays, i.e., decays of quarkonium states into photons, leptons, and gluons, sheds light on the perturbative aspect of QCD. Investigation of these decays is also helpful in the production, identification of conventional or non-conventional mesons, multiquark structures, etc. Using predicted masses and applying formulas obtained from Van Royen-Weisskopf formula, we estimate leptonic, digamma, tri gamma, digluon decay widths. These calculated decay widths are listed in Table \ref{Dilepton}, Table \ref{Digamma},  Table \ref{Trigamma}, Table \ref{Digluon}.
%%%%%%%%%%%%%%%%%%%%%%%%%%%%%%%%%%%%%%%%%%%%%%%%%%%%%%%%%%%%%%%%%%%%%%%%
\begin{table}[ht!] 

\caption{\label{Dilepton} Leptonic decay widths of bottomonia (in $keV$)}
\addtolength{\leftskip} {-2cm}
    \addtolength{\rightskip}{-2cm}
 \begin{tabular*}{1.2\linewidth}{@{\extracolsep{\fill}}ccccccccccc}\toprule
 $n^{2S+1}L_{j}$&$\Gamma_{l^+l^-}$& $\Gamma_{l^+l^-}^{C}$
  &Ref.\cite{pandya2021}
 & PDG\cite{PDG2022} & Ref.\cite{kher2022}& Ref. \cite{bhavsar2018}& Ref. \cite{wang2018}(eV)& Ref. \cite{segovia2016}\\
 \midrule
$ 1^{3}S_{1}$&0.91227& 0.59379 &0.7700&$1.340 \pm 0.018$&0.582&1.300&1.65&0.71\\
$ 2^{3}S_{1}$&0.65325& 0.42526 &0.5442&$0.612 \pm 0.011$&0.197&0.760&0.821&0.37\\
$ 3^{3}S_{1}$&0.57278& 0.37288 &0.4288&$0.443 \pm 0.008$&0.149&0.450&0.569&0.27\\
$ 4^{3}S_{1}$&0.48505& 0.34039 &0.3549&$0.272 \pm 0.029$&0.129&0.260&0.431&0.21\\
 $ 5^{3}S_{1}$&0.48505& 0.31578 &0.3035&$0.310 \pm 0.07$&0.117&0.180&0.348&0.18\\

$ 1^{3}D_{1}$ &10.96020 (eV)& 7.13748 (eV) &0.0050&-&1.65&0.106 (eV)&1.880 (eV)&1.40 (eV)\\ 
$ 2^{3}D_{1}$ &12.69930 (eV)& 8.26971 (eV) &0.0058&-&2.42&0.078 (eV)&2.810(eV)&2.50(eV)\\ 
$ 3^{3}D_{1}$ &18.73760 (eV)& 12.2004 (eV) &0.0059&-&3.19&0.051 (eV)&3.000(eV)&-\\ 
$ 4^{3}D_{1}$ &24.99260 (eV)& 16.2709 (eV) &0.0058&-&3.97&0.042(eV)&3.000(eV)&-\\ 
$ 5^{3}D_{1}$ &32.20100 (eV)& 20.9611 (eV) &0.0057&-&-&0.028 (eV)&0.003(eV)&-\\ 

 \bottomrule
\end{tabular*}
\end{table}
%%%%%%%%%%%%%%%%%%%%%%%%%%%%%%%%%%%%%%%%%%%%%%%%%%%%%%%%%%%%%%%%%%%%%%%%%%%%%%
\begin{table}[ht!]
\centering
\caption{ \label{Digamma} Digamma decay widths of $S$ and $P$ waves bottomonia (in $keV$) without and with QCD corrections}
\addtolength{\leftskip} {-2cm}
    \addtolength{\rightskip}{-2cm}
 \begin{tabular*}{1.2\linewidth}{@{\extracolsep{\fill}}ccccccccccc}\toprule
 $n^{2S+1}L_{j}$&$\Gamma_{\gamma \gamma}$& $\Gamma_{\gamma \gamma}^{C}$
  &Ref.\cite{nrsoni2018}
 & Ref.\cite{pandya2021} & Ref.\cite{godfrey2015}& Ref. \cite{kher2022}& Ref. \cite{wang2018}& Ref. \cite{li2009}\\
 \midrule
$ 1^{1}S_{0}$&0.30499& 0.23719 &0.3870&0.3035&0.940&0.2361&1.050&0.527\\
$ 2^{1}S_{0}$&0.24568& 0.19106 &0.2630&0.2122&0.410&0.0896&0.489&0.263\\
$ 3^{1}S_{0}$&0.23729& 0.18454 &0.2290&0.1668&0.290&0.0726&0.323&0.172\\
$ 4^{1}S_{0}$&0.23624& 0.18372 &0.2120&0.1378&0.200&0.0666&0.237&0.105\\
 $ 5^{1}S_{0}$&0.23739& 0.18462 &0.2010&0.1176&0.170&0.0636&0.192&0.121\\

$ 1^{3}P_{0}$&0.08412& 0.08522 &0.0196&0.1150&0.150&0.0168&0.199&0.037\\
$ 1^{3}P_{2}$ &0.07952& 0.01467 &0.0052&0.0147&0.093&0.0024&0.011&0.007\\
$ 2^{3}P_{0}$&0.07959& 0.08056 &0.0195&0.1014&0.150&0.0172&0.205&0.037\\
$2^{3}P_{2}$&0.08078& 0.01386 &0.0052&0.0131&0.012&0.0024&0.013&0.006\\
$ 3^{3}P_{0}$&0.08216& 0.08063 &0.0194&0.0875&0.130&0.0192&0.180&0.037\\
$3^{3}P_{2}$&0.02252& 0.01387 &0.0051&0.0114&0.013&0.0027&0.004&0.006\\
$4^{3}P_{0}$&0.02128& 0.08184 &0.0192&0.0768&0.130&-&0.157&-\\
$4^{3}P_{2}$&0.02129& 0.01408 &0.0051&0.0100&0.015&-&0.014&-\\
$ 5^{3}P_{0}$&0.02162& 0.08323 &0.0191&0.0686&-&-&0.146&-\\
$5^{3}P_{2}$&0.02198& 0.01431 &0.0050&0.0090&-&-&0.014&-\\
\bottomrule
\end{tabular*}
\end{table}
%%%%%%%%%%%%%%%%%%%%%%%%%%%%%%%%%%%%%%%%%%%%%%%%
\begin{table}[ht!]
\centering
\caption{\label{Trigamma} Trigamma decay widths of bottomonia (in unit of $10^{-3}$ $eV$) without QCD corrections}
\addtolength{\leftskip} {-2cm}
    \addtolength{\rightskip}{-2cm}
 \begin{tabular*}{1.2\linewidth}{@{\extracolsep{\fill}}cccccccc}\toprule
$n^{2S+1}L_{j}$ & $\Gamma_{\gamma \gamma \gamma}$
  &Ref.\cite{nrsoni2018}
 & Ref.\cite{wang2018} & Ref.\cite{godfrey2015}&Ref. \cite{kher2022}&Ref. \cite{segovia2016}& Ref.\cite{chaturvedi2020}\\
 \midrule
$ 1^{3}S_{1}$& 30.36900 &33.560&19.40&17.0&33.560&19.40&16\\
$ 2^{3}S_{1}$& 24.46270 &12.670&10.90&9.8&12.670&10.90&3\\
$ 3^{3}S_{1}$& 23.62710 &10.261&8.04&7.6&10.261&8.04&1\\
$ 4^{3}S_{1}$& 23.52240 &9.400&6.36&6.0&9.400&6.36&-\\
 $ 5^{3}S_{1}$& 23.63810 &8.979&5.43&-&8.979&5.43&-\\
\bottomrule
\end{tabular*}
\end{table}

\begin{table}[ht!] 

\caption{\label{Digluon} Digluon decay widths of bottomonia (in $MeV$) without and with QCD corrections}
\addtolength{\leftskip} {-2cm}
    \addtolength{\rightskip}{-2cm}
 \begin{tabular*}{1.2\linewidth}{@{\extracolsep{\fill}}ccccccccc}\toprule
$n^{2S+1}L_{j}$ & $\Gamma_{gg}$ & $\Gamma_{gg}^{C}$
  &Ref.\cite{nrsoni2018}
 & Ref.\cite{pandya2021} & Ref.\cite{godfrey2015}&Ref. \cite{kher2022}& Ref. \cite{wang2018}&Ref. \cite{segovia2016}\\
 \midrule
$ 1^{1}S_{0}$&4.33520& 5.58201 &5.448&6.8520&16.600&8.219&17.9&20.180\\
$ 2^{1}S_{0}$&3.49207& 4.49639 &3.710&5.2374&7.200&3.121&8.33&10.640\\
$ 3^{1}S_{0}$&3.37280& 4.34282 &3.229&4.3182&4.900&2.529&5.51&7.940\\
$ 4^{1}S_{0}$&3.35785& 4.32357 &2.985&3.6829&3.400&2.317&4.03&-\\
 $ 5^{1}S_{0}$&3.37436& 4.34483 &2.832&3.2196&2.900&2.214&3.26&-\\

$ 1^{3}P_{0}$&1.20151&  0.10870&0.276&1.4297&2.600&0.721&3.37&2.000\\
$ 1^{3}P_{2}$ & 0.32040&0.31831&0.073&0.2370&0.147&0.192&0.165&0.084\\
$ 2^{3}P_{0}$& 1.13537 &0.08409&0.275&1.2358&2.600&0.741&3.52&2.370\\
$2^{3}P_{2}$& 0.30276 &0.30078&0.073&0.2064&0.207&0.198&0.220&0.104\\
$ 3^{3}P_{0}$& 1.13611 &0.07036&0.273&1.0539&2.200&0.828&3.10&2.460\\
$3^{3}P_{2}$& 0.30296 &0.30098&0.072&0.1767&0.227&0.221&0.243&0.112\\
$4^{3}P_{0}$& 1.15308 &0.06057&0.271&0.9175&2.100&-&-&-\\
$4^{3}P_{2}$& 0.30749& 0.30547&0.072&0.1543&0.248&-&-&-\\
$ 5^{3}P_{0}$& 1.17255 &0.05283&0.269&0.8127&-&-&-&-\\
$5^{3}P_{2}$& 0.31268 &0.31063&0.071&0.1370&-&-&-&-\\
$ 1^{1}D_{2}$ & 5.93167 ($keV$) &-&-&-&1.800 ($keV$)&0.489 ($keV$)&0.657 ($keV$)&0.370 ($keV$)\\ 
$ 2^{1}D_{2}$ & 7.53656 ($keV$) &-&-&-&1.530 ($keV$)&0.764 ($keV$)&1.22 ($keV$)&0.670 ($keV$)\\ 
$ 3^{1}D_{2}$ & 12.10030 ($keV$) &-&-&-&1.839 ($keV$)&1.0006 ($keV$)&1.59 ($keV$)&-\\ 
$ 4^{1}D_{2}$ & 17.45940 ($keV$) &-&-&-&-&1.380 ($keV$)&1.86 ($keV$)&-\\ 
$ 5^{1}D_{2}$ & 24.21910 ($keV$) &-&-&-&-&-&2.13 ($keV$)&-\\ 

 \bottomrule
\end{tabular*}
\end{table}
 
On observing Table \ref{Digamma}, we found that calculated partial decay widths for $nS$ states agree with other models' predictions.
But our results for states $nP$ are slightly off from other models' predictions. They differ from other predictions by $\pm 65$ $eV$. We also calculated tri gamma decay widths mentioned in Table \ref{Trigamma}, and the results are in good accord with estimations deduced by other theoretical approaches.

The calculated digluon decay widths are listed in Table \ref{Digluon} and compared with predictions of other models. On comparing, it is found that our estimations are consistent with the results of other models for $nS$ states. However, our results for $nP$ states are higher than the predictions of other models. A similar trend can be seen in the case of $nD$ states, i.e., our results are slightly higher than other model estimations. 

The dileptonic decay widths of quarkonium play an important role in the estimation of strong coupling constants,  decay constants and in checking the validity of theoretical models as its decay amplitude carries quarkonium wave function (can be seen in eqn.\ref{leptoniceq1}). We calculated leptonic decay width and mentioned it in Table \ref{Dilepton}. On comparing, it is found that our predictions for $nS$ states agree with the outcomes of other models. For $nD$ states, our results agree with Ref. \cite{pandya2021} and are higher in comparison to Ref. \cite{bhavsar2018, wang2018,segovia2016}. For higher $n$, our decay widths for $D$ states show a monotonically increasing trend, which agrees with some of the compared models. It is also noticed on observing tables of annihilation decays that for $D$ wave states, values of leptonic decays and digluon decays are highly suppressed as compared to $S$ waves, which is expected.

As we know, the total decay width of a particular state is the sum of strong, radiative, and weak decay widths. But strong decay gives more contribution to the total decay width of that state relative to radiative and weak decay. This is inferred from Table \ref{E1}, \ref{M1}, \ref{Dilepton}, \ref{Digamma}, \ref{Trigamma}, \ref{Digluon}, which reflects the value of digluon decays (strong decays) are much more relative to radiative transitions and leptonic decays (weak decays).
\section{Conculsion}
In this paper, we have presented a comprehensive analysis of the mass spectra of bottomonium states and their decay properties by employing a non-relativistic potential model. Using adopted quark-antiquark confining potential, we solved Schrodinger wave equations numerically with Runge -Kutta method using the Mathematica notebook \cite{lucha1999}. Our calculated mass spectrum ($nS, nP, nD, nF, n = 1, 2, 3, 4, 5$) of bottomonia is fairly close to available experimental data and other theoretical model predictions. The hyperfine splittings of $1P$ and $2P$ states are in fair agreement with the experiment observation. We have computed annihilation decays - di-leptonic, di-gamma, tri-gamma, di-gluons decay widths. Also, the electromagnetic transition widths by Van Royen - Weisskopf formula are calculated. All these decays are also estimated with first-order QCD corrections. Our findings agree with the available experimental observations and theoretical estimations. We have also explored leptonic decay constants, which are consistent with the experimental values mentioned in PDG. The present results may be helpful in upcoming experimental information in the near future.     
\section{Acknowledgement}
The authors gratefully acknowledge the financial support by the
Department of Science and Technology (SERB/F/9119/2020), New
Delhi and for Junior Research Fellowship (09/0677(11306)/2021-EMR-I) by Council of Scientific and Industrial Research, New Delhi. 

\bibliography{ref}
\bibliographystyle{epj}
\end{document}